\documentclass[letterpaper, 10 pt, conference]{ieeeconf}  

\IEEEoverridecommandlockouts                              

\usepackage{xcolor} 
\definecolor{mygreen}{RGB}{130,179,102}
\definecolor{myred}{RGB}{183,84,79}

\usepackage{graphicx}
\usepackage{mathptmx}
\usepackage{amsmath}
\usepackage{amssymb}

\usepackage{amsthm}
\usepackage[english]{babel}
\usepackage{csquotes}

\newtheorem{remark}{Remark}
\newtheorem{lemma}{Lemma}

\usepackage[
backend=biber,
style=ieee,
sorting=none,
doi=false,isbn=false,url=false 
]{biblatex}
\usepackage{algorithm2e}
\usepackage{multirow}
\usepackage{booktabs}

\title{
Towards Optimal Kron-based Reduction Of Networks (Opti-KRON) \\ for the Electric Power Grid}

\author{Samuel Chevalier$^{1}$ and Mads R. Almassalkhi$^{2}$
\thanks{$^{1}$Samuel Chevalier is a Marie Curie Postdoctoral Fellow in the new Department of Wind and Energy Systems at Denmark's Technical University (DTU), Kongens Lyngby, Denmark, and acknowledges the HORIZON-MSCA-2021 Postdoctoral Fellowship Program, Project \#101066991 -- TRUST-ML. Email: {\tt\small schev@elektro.dtu.dk}}%
\thanks{$^{2}$ Mads R. Almassalkhi is the L. Richard Fisher Associate Professor in Electrical Engineering at the University of Vermont (UVM) in Burlington, VT, Chief Scientist at Pacific Northwest National Laboratory (PNNL), in Richland, WA,  and Otto M\o nsted Visiting Professor in the Department of Wind and Energy Systems at Denmark's Technical University (DTU), Kongens Lyngby, Denmark, and acknowledges support from the U.S. National Science Foundation (NSF) Award ECCS-2047306 and the Otto M\o nsted Fund. {Email: \tt\small malmassa@uvm.edu}}
}

\begin{document}

\maketitle
\thispagestyle{empty}
\pagestyle{empty}

\begin{abstract}

For fast timescales or long prediction horizons, the AC optimal power flow (OPF) problem becomes a computational challenge for large-scale, realistic AC networks. To overcome this challenge, this paper presents a novel network reduction methodology that leverages an efficient mixed-integer linear programming (MILP) formulation of a Kron-based reduction that is optimal in the sense that it balances the degree of the reduction with resulting modeling errors in the reduced network. The method takes as inputs the full AC network and a pre-computed library of AC load flow data and uses the graph Laplacian to constraint nodal reductions to only be feasible for neighbors of non-reduced nodes. This results in a highly effective MILP formulation which is embedded within an iterative scheme to successively improve the Kron-based network reduction until convergence. The resulting optimal network reduction is, thus, grounded in the physics of the full network. The accuracy of the network reduction methodology is then explored for a 100+ node medium-voltage radial distribution feeder example across a wide range of operating conditions. It is finally shown that a network reduction of 25-85\% can be achieved within seconds and with worst-case voltage magnitude deviation errors within any super node cluster of less than 0.01pu. These results illustrate that the proposed optimization-based approach to Kron reduction of networks is viable for larger networks and suitable for use within various power system applications.

\end{abstract}

\section{INTRODUCTION}

Understanding how to best utilize resources distributed over a network has been and is an important question across many industries. For the power/energy industry, solving the centralized AC optimal power flow (OPF) problem is NP-hard and has been the focus of much research since the 1960s~\cite{squires_economic_1960,dommel_optimal_1968,carpentier_optimal_1979} and more recently, as optimization solvers matured~\cite{molzahn_survey_2019}. In some cases, the OPF problem is cast within the setting of (transmission) expansion planning and considers a large number of scenarios, decade-long prediction horizons, and many possible investment decisions~\cite{ploussard_efficient_2018}. In other cases, the focus of the OPF problem is near-term grid operations to determine active and/or reactive power set-points for PV inverters, batteries, and other controllable assets in the grid to minimize operating costs, line losses, voltage deviations from nominal, or to achieve a desired net-load profile that reflects whole-sale energy market conditions~\cite{almassalkhi_hierarchical_2020}.
Thus, many applications of the AC OPF requires a mix of long prediction horizons and frequent re-computations, which for large networks are computationally challenging. 

To overcome the computational challenges associated with solving practical (large-scale) AC OPF problems to (global) optimality, the power/energy community has often studied approximations of the AC physics, such as the so-called (linear) DC power flow~\cite{stott_dc_2009}, convex relaxations~\cite{low_convex_2014}, convex restrictions~\cite{dongchan_robust_2021}, and various distributed implementations~\cite{dallanese_optimal_2018}. However, if the AC network could be reduced, while representing the physics of the network sufficiently well, the computational challenge would decrease significantly~\cite{rogers_aggregation_1991}. Thus, in this paper, we focus on a novel method for (optimally) reducing the AC network, which could then be employed within an appropriate OPF setting. 

Network reductions, not to be confused with model-order reduction from systems theory~\cite{kokotovic_singular_1976,chevalier_accelerate_2021}, have been studied extensively and employ a variety of methods, such as similarity or (electrical) distance measures for clustering, bus aggregations (e.g., REI), and equivalence techniques (e.g., Ward and Kron). In the case of reducing nodes belonging to an ``external'' area, which are nodes that are geographically or electrically distanced from the ``internal'' area, network reduction via Ward- or Kron-based methods can be readily applied and has been standard practice for decades~\cite{ploussard_efficient_2019}. However, recently techniques have focused inwards on the internal area or so-called ``backbone-type'' network reductions, where any nodes can be reduced in the network rather than just ``external'' nodes. These backbone-type equivalents rely on either an initial clustering approach (e.g., $k$-means clustering) to group nodes together into contiguous zones or a pre-defined set of zones. Once the nodes are assigned to specified zones (or subgrids), a network reduction can be readily applied to said zones (e.g., via Ward and Kron or heuristics) and possibly tuned based on some criteria.  For example, network-preserving bus aggregation methods by ~\cite{fortenbacher_transmission_2018} and~\cite{shi_novel_2015} employ nonlinear and quadratic optimization, respectively, to tune (susceptance values in) the reduced admittance matrix so as to minimize tie line flow errors with respect to the full network. In~\cite{fortenbacher_transmission_2018}, the method depends on pre-determined zones and a specific operating point to calculate the full network's power transfer distribution factors (PTDFs). The algorithm~\cite{shi_novel_2015} replaces the zonal input requirement with a list of pre-determined salient tie lines and also uses PTDFs, which inform a bus clustering algorithm that defines internal zones, which are then subject to bus aggregation. These methods can reduce 60,000-bus networks by up to 100X in the order of minutes (on a super computer) with small inter-zonal worst-case flow deviation errors - even under different operating conditions. Other approaches sidestep the dependence on operating points by employing DC load flow analysis in deriving independent PTDF values~\cite{oh_new_2010}. In this case, a 15,000-bus network is reduced by 85X after eight hours with relative line flow errors of less than 30\%. Lastly, some methods are built around multiple clustering objectives and heuristics that preserve physical features and network structure, but are overly conservative (i.e., only reduce by 2-3X while line flow deviation errors are around 5-10\%)~\cite{sistermanns_feature-_2019}. 

Kron-based network reductions have been shown to be valuable across numerous applications in power system analysis~\cite{dorfler_kron_2013,ploussard_efficient_2019}. For example, comprehensive transmission planning schemes have been built around Kron-based equivalents that employ various optimization formulations whose solutions serve as seeds to identify a set of salient buses and lines to partition the network~\cite{ploussard_efficient_2019}. To speed up 3-phase distribution grid OPF,~\cite{almassalkhi_hierarchical_2020} presents a Kron-based network reduction, where a desired level of reduction informs a nodal clustering scheme that determines which nodes are reduced. This method was able to achieve 10-50X reductions in realistic distribution feeders with maximum voltage deviation errors (between reduced nodes and their corresponding non-reduced ``super'' node in the same cluster) of less than 0.015pu across a wide range of operating conditions.

Across these different approaches to network reductions in power systems, they all depend on pre-specified salient buses, tie-lines, and/or level of desired reduction as inputs. Clearly, these inputs affect the resulting network reduction and this is what motivates a simple, but interesting question: \textit{is there an optimal Kron-reduction?} More precisely: are there a set of nodes and a level of reduction that is optimal (in some sense) when reducing a network? Thus, as a first step towards answering this question, the paper's key contribution is a novel network reduction methodology that leverages a mixed-integer linear programming (MILP) formulation to determine a Kron-based reduction that is optimal in the sense that it automatically balances the level of reduction (i.e., complexity) with resulting worst-case voltage deviation errors between the reduced and full networks. The method is based on a pre-computed library of AC load flow data (i.e., operating points) and guarantees that any feasible solution is a valid Kron reduction that preserves the network's structure. As far as the authors are aware, there is no other literature that casts a Kron-based network reduction entirely within an efficient MILP optimization formulation. To ensure tractability in the MILP formulation, we constrain nodes to only reduce to a ``super node,'' if they are neighbors (as defined by the graph Laplacian). Then, we successively reduce the network via an iterative scheme to overcome the nodal neighbor limitation. The entire methodology, denoted Opti-KRON, is validated via simulation-based analysis on a 115-node, radial and balanced IEEE test network, which represents a minor contribution as it provides insight on different optimal Kron-based network reductions. The algorithm is also tested in the context of a 200-node mesh network.

The remaining paper is structured as follows. Section~\ref{sec:model} presents the network model and summarizes Kron reductions. In Section~\ref{sec:method}, the MILP formulation for Kron-based network reduction is presented. Simulation-based analysis is presented in Section~\ref{sec:exampleLarger}. Finally, the paper concludes in Section~\ref{sec:end} with a summary and a brief discussion on future directions and applications.

\section{Network model and Kron-reduction}\label{sec:model}
For the sake of notational simplicity, consider a single-phase power system network whose graph ${\mathcal G}({\mathcal V},{\mathcal E})$ has edge set $\mathcal{E}$, $|\mathcal{E}|=m$, vertex\footnote{In power systems, a vertex in a power network is commonly denoted a node in distribution systems and a busbar (or bus) in transmission systems. Given the general discussion of power networks, we will use node and bus interchangeably.} set $\mathcal{V}$, $|\mathcal{V}|=n$, and signed nodal incidence matrix $E\in{\mathbb R}^{m\times n}$. The complex nodal admittance matrix (i.e., Y-bus matrix) ${Y}_b\in{\mathbb C}^{n\times n}$ associated with this system is constructed via
\begin{equation}\label{eq: Yb}
{Y}_{b}=E^{T}{Y}_{l}E+{Y}_{s},
\end{equation}
where ${Y}_{l}\in{\mathbb C}^{m\times m}$ is the diagonal matrix of complex line admittances and ${Y}_{s}\in{\mathbb C}^{n\times n}$ is the diagonal matrix of complex nodal shunt admittances. In this paper, we generally assume ${Y}_{s}\ne 0$, implying ${Y}_{b}$ is a nonsingular matrix. Leveraging this property, the so-called nodal impedance matrix ${ Z}_b\in{\mathbb C}^{n\times n}$ can be directly computed as the inverse of (\ref{eq: Yb}): ${Z}_{b}={Y}_{b}^{-1}$. The nodal admittance (and impedance) matrices directly relate complex nodal voltages and current injections via ${I}={Y}_{b}{V}$ (and ${Z}_{b}{I}={V}$).

\subsection{The Kron-Reduction Procedure}\label{sec:method}
Without loss of generality, we partition the network via
\begin{align}\label{eq: Y_partition}
{I} & ={Y}_{b}{V}\\
\left[\begin{array}{c}
{I}_{k}\\
\hline  {I}_{r}
\end{array}\right] & =\left[\begin{array}{c|c}
{Y}_{b1} & {Y}_{b2}\\
\hline {Y}_{b3} & {Y}_{b4}
\end{array}\right]\left[\begin{array}{c}
{V}_{k}\\
\hline {V}_{r}
\end{array}\right],
\end{align}
where subscripts ``$r$" and ``$k$" denote nodes which are ultimately reduced and kept, respectively. As in~\cite{dorfler_kron_2013}, Gaussian elimination of the nodal voltages ${ V}_r$ is achieved by
\begin{align}
{I}_{k}=\left({Y}_{b1}-{Y}_{b2}{Y}_{b4}^{-1}{Y}_{b3}\right){V}_{k}+\left({Y}_{b2}{Y}_{b4}^{-1}\right){I}_{r}.
\end{align}
The Kron reduction of (\ref{eq: Y_partition}), which is used to ``eliminate" nodes with zero current injection (i.e., $I_r=0$), is canonically given by the following Schur complement:
\begin{align}
{Y}_{K}={Y}_{b1}-{Y}_{b2}{Y}_{b4}^{-1}{Y}_{b3}.
\end{align}
Alternatively, ${Y}_{K}$ can be constructed using the network impedance matrix, whose associated partition is given by
\begin{align}\label{eq: Zb}
\left[\begin{array}{c|c}
{Z}_{b1} & {Z}_{b2}\\
\hline {Z}_{b3} & {Z}_{b4}
\end{array}\right]\left[\begin{array}{c}
{I}_{k}\\
\hline 0
\end{array}\right]=\left[\begin{array}{c}
{V}_{k}\\
\hline {V}_{r}
\end{array}\right].
\end{align}

\begin{remark}
The Kron-reduced admittance ${Y}_{K}$ is equal to the inverse of sub-impedance ${Z}_{b1}$.
\end{remark} \begin{proof} By construction, the Kron admittance relates ${I}_{k}={Y}_{K}{V}_{k}$. From (\ref{eq: Zb}), ${Z}_{b1}{I}_{k}={V}_{k}$. Therefore, ${Z}_{b1}={Y}_{K}^{-1}$.
\end{proof}

\begin{figure}[t]
\includegraphics[width=\columnwidth]{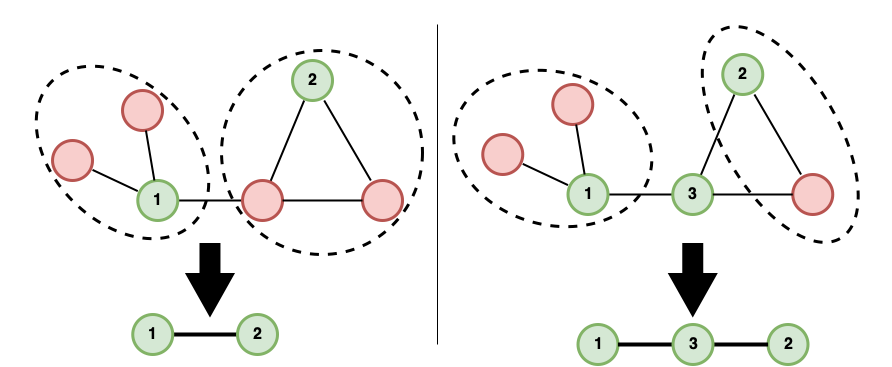}
\caption{Illustration of how a network can be partitioned in two different ways to yield two different Kron-reduced networks. The partition is based on reduced and kept (super) nodes. The {\color{mygreen}green}, numbered circles represent kept nodes or super nodes, while {\color{myred}red} circles are reduced nodes and eliminated in the reduced network. The dashed (\textbf{-\,-\,-}) ellipses illustrate which reduced nodes are assigned to which super nodes and define how injected currents are assigned to each super node.}
\centering
\label{fig:explainKron}
\end{figure}

In the following, we define the Kron impedance matrix as ${Z}_{K}\triangleq{Z}_{b1}$
from (\ref{eq: Zb}). Note that any removal of rows and columns from ${Z}_{K}$ will result in a valid Kron impedance matrix, in the sense that it will relate nodal voltages and currents. {Thus, if we can optimally select which nodes to reduce and where to assign them, we can effectively choose an optimal set of rows and columns to remove from ${Z}_{K}$. This would then allow us to define an optimal Kron impedance matrix through a set of binary decisions, which is illustrated in Fig.~\ref{fig:explainKron}.} This inspires the following mixed-integer formulation.

\subsection{Mixed-Integer Approach for Constructing Kron Matrices}\label{sec:modelMIP}

In the following, we define binary variable $s_i\in\{0,1\}$, which selects the optimal Kron impedance matrix $Z_k$, where $s_i=0$ or $1$ indicates if the $i^{\rm th}$ node is reduced or kept, respectively. We accordingly define binary vector $s\in \{0,1\}^{n}$ and the associated diagonal selection matrix $S\triangleq{\rm diag}\{s\}$. For any given binary values, the matrix product $S{Z}_{b}S$ thus yields a matrix which we refer to as a generalized Kron impedance, defined as ${Z}_{\hat{K}}\triangleq S{Z}_{b}S$. This generalized Kron impedance has the dimensions of the full nodal impedance matrix ($n\times n$), but a subset of its rows and columns are zeroed-out. As an example, the generalized Kron impedance of (\ref{eq: Zb}) is
\begin{align}
{Z}_{\hat{K}} & =\left[\begin{array}{c|c}
I & 0\\
\hline 0 & 0
\end{array}\right]\!\!\left[\begin{array}{c|c}
{Z}_{b1} & {Z}_{b2}\\
\hline {Z}_{b3} & {Z}_{b4}
\end{array}\right]\!\!\left[\begin{array}{c|c}
I & 0\\
\hline 0 & 0
\end{array}\right]\!=\!\left[\begin{array}{c|c}
{Z}_{b1} & 0\\
\hline 0 & 0
\end{array}\right],
\end{align}
where the diagonal binary values of $S$ ``kept" the top nodes and ``reduced" the bottom ones.

We define an additional binary decision matrix, $A\in \{0,1\}^{n \times n}$, which codifies where currents from reduced nodes are placed. Accordingly, $A_{i,j}=1$ if the nodal current injection from bus $j$ is placed at bus $i$, and $A_{i,j}=0$ otherwise. Since current can only be assigned to a single bus, $\sum_i A_{i,j}=1$ is always enforced. Furthermore, $\sum_j A_{i,j}\le M_b S_{i,i}$ ensures that currents cannot be assigned to a reduced bus and $S_{i,i} = A_{i,i}$ guarantees that each non-reduced bus does not move its own current. Based on these rules, the matrix vector product ${ I}_K = A{ I}$ naturally and properly aggregates currents at non-reduced nodes (i.e., Kron currents), and the following product allows for the direct computation of Kron voltages:
\begin{align}\label{eq: Vk}
    {V}_{K}=S{Z}_{b}SA{I}.
\end{align}
We define the non-reduced nodes as ``super nodes", and the Kron reduced voltages at these super nodes are given by (\ref{eq: Vk}).

In order to compute an \textit{optimal} Kron reduction, we need to define an objective function, which balances the trade off between complexity (i.e., level of detail) and corresponding nodal voltage deviation error (i.e., performance of reduced network). As the number of reduced nodes is a measure of reduction in network complexity, we can capture this by minimizing the number non-zero binary values in the reduction vector $s$. In order to quantify voltage deviation error (which generally increases as network reduction increases), we take the infinity norm of the difference between the Kron voltage at the $i^{\rm th}$ super node (${V}_{K,i}$) and the voltages at all nodes within its cluster $C_{K,i}$, across all potential super nodes. The resulting objective function is then given by
\begin{align}\label{eq: Loss}
\mathcal{L}=\underbrace{\left\Vert {V}_{K,i}-{V}_{j\in C_{K,i}}\right\Vert_{\infty}}_{{\rm error}} - \alpha \underbrace{\frac{1}{n}\sum_{j=1}^n(1-s_{j})}_{{\rm complexity}},
\end{align}
where $\alpha$ balances these two terms. We note that the network currents ${I}$ in (\ref{eq: Vk}) and the cluster voltages ${V}_{j}$ in (\ref{eq: Loss}) are assumed to be given as input data libraries for the optimization problem. Ideally, these data vectors (or matrices) come from representative AC power flow solutions collected on the full network. Thus, an optimal (with respect to (\ref{eq: Loss})), yet, naive binary integer-based Kron reduction can be stated as
\begin{subequations}\label{eq: Kron}
\begin{align}
\min_{s,A}\;\; & \left\Vert {V}_{K,i}-{V}_{j\in C_{K,i}}\right\Vert _{\infty}-\frac{\alpha}{n}\sum_{j=1}^{n}(1-s_{j})\label{eq: obj}\\
{\rm s.t.}\;\; & {V}_{K}=S{Z}_{b}SA{I}\label{eq: SZS_opt}\\
 & s_{i},\;A_{i,j}\in\{0,1\}\\
 & S={\rm diag}(s)\\
 & \ensuremath{\ensuremath{\sum_{i}A_{i,j}=1}}\label{eq: Aij1}\\
 & \ensuremath{\sum_{j}A_{i,j}\le M_{b}S_{i,i}}\label{eq: Aij}\\
 & \ensuremath{S_{i,i}=A_{i,i}}\label{eq: Aijs}.
\end{align}
\end{subequations}
While (\ref{eq: Kron}) will generally compute a valid Kron impedance matrix in (\ref{eq: SZS_opt}), the given formulation presents a variety of challenges. First, it does not formally constrain current injections associated with reduced nodes from being placed on super nodes which are electrically or geographically ``far" from their physical location. Second, the product $S Z_b S AI$ in (\ref{eq: SZS_opt}) contains cubic binary terms. And third, this problem is generally intractable for large-scale systems, since matrix $A$ is a binary matrix which engenders a large branch-and-bound search space for MILP solvers. In the following subsection, we address all three of these challenges to engender a tractable MILP-based Kron reduction.

\subsection{Formulation Improvements  }\label{sec:modelMIP_update}

The issue of cubic binary terms in (\ref{eq: SZS_opt}) is sidestepped by first simplifying the product term $SA$.
\begin{lemma}
$SA=A$.
\begin{proof}
Constraint (\ref{eq: Aij}) forces the $j$-th row of $A$ to 0 if $S_{j,j} = 0$; thus, reduced nodes (i) must place their currents somewhere else ($A_{j,j}=0$) and (ii) cannot receive currents from other reduced nodes ($A_{j,i}=0$). If when a binary, whose value is 0, only multiplies other binaries whose value is also 0, then the original binary has no effect. Thus, $SA=A$.
\end{proof}
\end{lemma}

Using $SA=A$, we now have that ${V}_{K}=S{Z}_{b}A{I}$. However, we cannot apply the same trick again to simplify $SZ_{b}A$ since $Z_{b}$ is generally a dense matrix. This means that each element of $A$ is multiplied by each diagonal element of $S$. Since the product of any two binaries can be reformulated in linear form (thus, ``linearizing'' the expression) by introducing a third auxiliary binary variable, directly linearizing $SZ_{b}A$ will generally require $n^3$ binary auxiliary variables; e.g., binary matrices $B_1 = S_{1,1}\times A$, $B_2= S_{2,2}\times A$, etc. Rather than directly linearizing this expression, however, we can leverage the physically motivated observation that any error accumulated by removing $S$ from the Kron voltage equation (denoted with a tilde: $\tilde{V}_{K}={Z}_{b}A{I}$) can be subsumed into auxiliary big-M slack factors. To do so, we reformulate the infinity norm in the objective function of (\ref{eq: Kron}) with a continuous slack variable $\delta$:
\begin{subequations}\label{eq: Kron_simp}
\begin{align}\min_{s,A,\delta}\;\; & \delta-\frac{\alpha}{n}\sum_{j=1}^{n}(1-s_{j})\\
{\rm s.t.}\;\; & \tilde{V}_{K,i}-V_{j\in C_{K,i}}\le\delta+M_{b}(1-A_{i,j}),\forall i\label{eq: vkt}\\
 & V_{j\in C_{K,i}}-\tilde{V}_{K,i}\le\delta+M_{b}(1-A_{i,j}),\forall i\label{eq: -vkt}\\
 & \tilde{V}_{K}=Z_{b}AI\label{eq: Vktil}\\
 & \eqref{eq: Aij1}-\eqref{eq: Aijs}\nonumber
\end{align}
\end{subequations}

\begin{lemma}
Despite the Kron voltage error in (\ref{eq: Vktil}) caused by the elimination of $S$, (\ref{eq: Kron_simp}) and (\ref{eq: Kron}) have identical minimizers.

\begin{proof}
Multiplying $S{\tilde V}_K$ yields ${V}_K$, so $S$ effectively zeros-out non-super node voltages. However, it does not change the value of the super node voltage itself. $A_{i,j}=1$ indicates that node $j$ is inside the cluster associated with super node $i$. In this case, $M_{b}(1-A_{i,j})=0$, and $\delta$ will be a supremum for the exact intra-cluster voltage deviations (since super node voltages are preserved in ${\tilde V}_K$). However, when $A_{i,j}=0$, node $j$ is not internal to the cluster associated with node $i$, which may or may not be a super node. Therefore, $M_{b}(1-A_{i,j})=M_{b}$ will safely upper bound any voltage deviation between ${\tilde V}_K$ and $V_{j\in C_{K,i}}$, thus leaving $\delta$ unaffected. Since $\delta$ accurately captures the infinity norm value from (\ref{eq: obj}), the programs must have identical minimizers.
\end{proof}
\end{lemma}

In order to avoid allowing the optimizer to add current from reduced nodes far from the super node itself, we employ the graph Laplacian to constrain current aggregation only at neighboring nodes.
To accomplish this, we enforce the binary values in matrix $A$ (which chooses where currents are aggregated) to satisfy 
\begin{align}\label{eq: Laplacian}
A_{i,j}\le|E^{T}E|_{i,j},
\end{align}
where $|E^{T}E|_{i,j}$ is the $i,j$-th entry of the absolute value of the graph Laplacian. Therefore, if two nodes are not direct neighbors, then their currents cannot be aggregated together. Not only does this prevent currents from being placed in non-physically meaningful places, it also greatly limits the size of the search space, thus greatly increasing the tractability of (\ref{eq: Kron}). 

\subsection{Successive Enhancement of Reduced Networks }\label{sec:modelMIP_iterate}
\begin{figure}[t]
\includegraphics[width=\columnwidth]{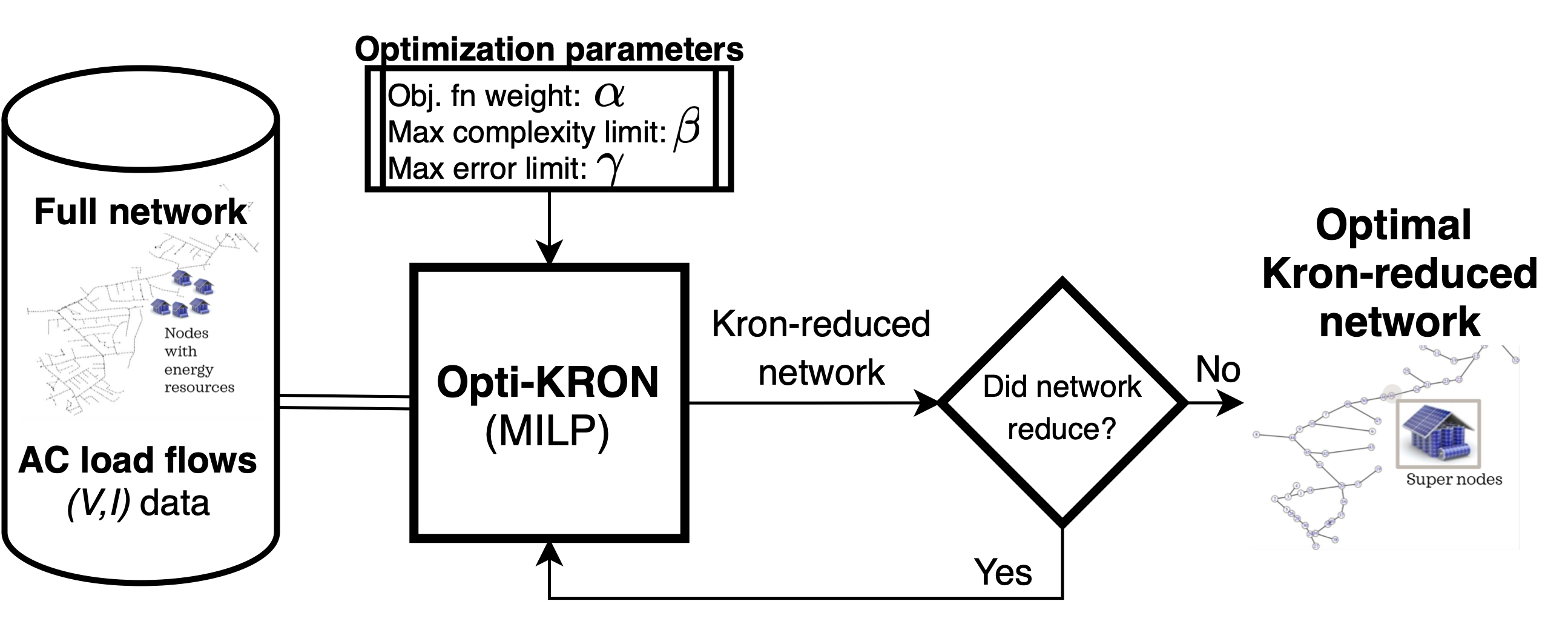}
\caption{The algorithm for successively enhancing the Kron-reduced network uses Opti-KRON, which is given by~\eqref{eq:optiKron}. The inputs are network and AC load flow data and the parameters that define the MILP formulation's objective function. The output is an optimal Kron-reduced network where kept nodes are denoted super nodes.}
\centering
\label{fig:algoIter}
\end{figure}

While (\ref{eq: Kron_simp}) \& (\ref{eq: Laplacian}) jointly represent a highly tractable mathematical program, the degree of reduction it can achieve is limited by the graph Laplacian constraint in~\eqref{eq: Laplacian}. Since nodes can only aggregate with their neighbors, the algorithm cannot typically achieve more than a 60\% network reduction in a given solve. To overcome this hurdle, we propose an iterative implementation. That is, find an optimal network reduction, construct the reduced network, and then find another optimal network reduction of the pre-reduced network. This procedure is repeated until either (i) the desired level of reduction is achieved, or (ii), zero nodes are reduced. In order to control the maximal size $\beta$ of network reduction (i.e., force optimizer to make small network reductions at each step), or control the maximal acceptable voltage deviations $\gamma$, we can embed associated constraints directly in the program.
\begin{subequations} \label{eq:optiKron}
\begin{align}\min_{A,\delta}\;\; & \delta-\frac{\alpha}{n}\sum_{j=1}^{n}(1-A_{j,j})\\
{\rm s.t.}\;\; & \sum_{j=1}^{n}(1-A_{j,j})\le n\beta\\
 & \delta\le\gamma\\
 &  \eqref{eq: Aij1}-\eqref{eq: Aijs},\eqref{eq: vkt} - \eqref{eq: Vktil},\eqref{eq: Laplacian}\nonumber.
\end{align}
\end{subequations}
This iterative approach is illustrated in  Fig.~\ref{fig:algoIter} and described algorithmically in Algorithm~\ref{alg:one}. We refer to the tractable mathematical program given by~\eqref{eq:optiKron} as Opti-KRON. Next, we apply Opti-KRON and Algorithm~\ref{alg:one} to optimally Kron reduce an IEEE test network, which represents a balanced, medium-voltage, radial distribution feeder.

\RestyleAlgo{ruled}
\SetKwComment{Comment}{/* }{ */}
\begin{algorithm}
\caption{Opti-KRON Successive Enhancement }\label{alg:one}
\KwData{$Y_b, V, I, \alpha, \beta, \gamma$}
\KwResult{$Z_K$ (Optimal Kron-reduced network)}
$p=0, s^{(p)}=\mathbf{1}_n,  \Delta s^{(p)}=n$ \;
${Z}_{b} = {Y}_{b}^{-1}$ \;
 \While{$\Delta s^{(p)}>0$}{
  $s^{(p+1)} \gets$ Solve Opti-KRON in \eqref{eq:optiKron}\;
  $\Delta s^{(p+1)} =  \mathbf{1}_n^\top (s^{(p)} - s^{(p+1)})$ \;
  $p \gets p+1$ \;
 }
 $S \gets \text{diag}\{s^{(p)}\}$ \,\&\, $Z_K \gets S Z_b S A$ \;
\end{algorithm}


\section{Experimental results} \label{sec:exampleLarger}
In this section, we provide test results collected from two systems: a 115-node radial network, and a 200-node radial network.

\subsection{115-node radial network}
\begin{figure}[t]
\includegraphics[width=\columnwidth]{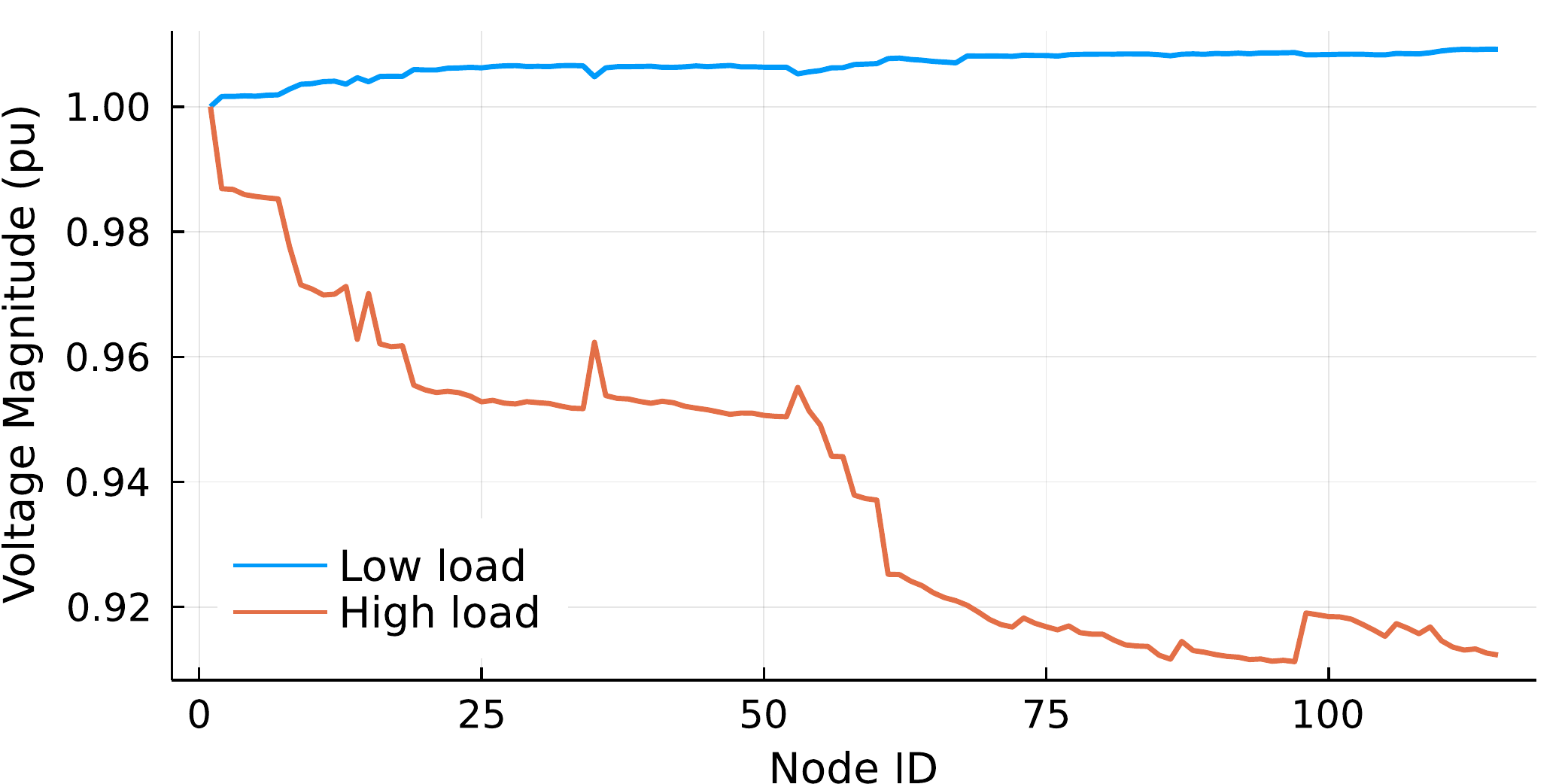}
\caption{The voltage profile resulting from two distinct net-load injections in the radial 115-node network. The red line represents a heavily loaded scenario, while the blue line represents a lightly loaded scenario with more solar PV injections.}
\label{fig:voltProfiles}
\centering
\end{figure}

The 115-node radial network represents a single phase from the IEEE 123-node distribution test feeder~\cite{schneider_ieee_2017}. This network provides $Y_b$ and is used herein to illustrate Algorithm~\ref{alg:one}. To balance complexity and error, $\alpha= 0.002$, while the maximum reduction in complexity for a single iteration of~\eqref{eq: Kron_simp} is limited initially to 25\% (i.e., $\beta = 0.25$). The worst-case voltage deviation error is effectively unconstrained by setting $\gamma = 1.0$pu. Finally, two distinct nodal net-injection profiles are applied to the network to beget the network's necessary data scenarios on complex branch currents, $I$, and nodal voltages, $V$. The corresponding voltage profiles, $|V_i| \hspace{1mm} \forall i=1,\hdots, n$, are shown in Fig.~\ref{fig:voltProfiles}.

\begin{figure}[t]
\includegraphics[width=\columnwidth]{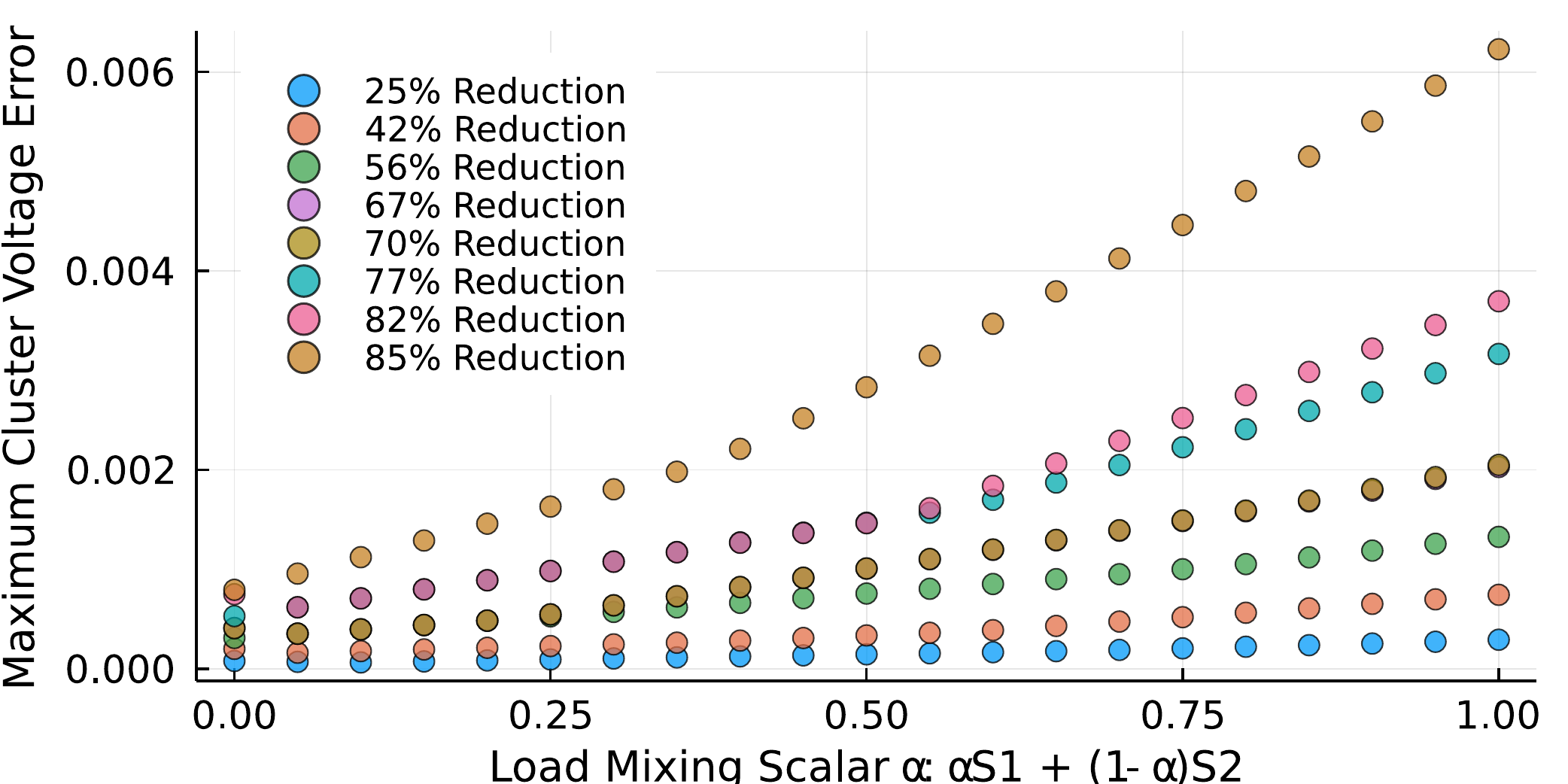}
\caption{Worst-case intra-cluster voltage errors for all iterate Kron-reduced versions of the radial 115-node network (optimally reduced via eight iterations in Algorithm~\ref{alg:one}). 
}
\label{fig:volterrorsIters}
\centering
\end{figure}

With all input data now available, Algorithm~\ref{alg:one} can be executed and converges in eight iterations and under five seconds total, which highlights tractability of Opti-KRON. The resulting optimal Kron-based network reduction has eliminated 85\% of nodes, yet embodies a worst-case intra-cluster voltage deviation error across both load scenarios of less than $0.007$pu. To investigate the accuracy of Opti-KRON, we subject the optimal Kron reduction \textit{at each iteration} to operating conditions that sweep from low-load to high-load conditions (via a convex combination of the initial injection data). Then, we record the maximum intra-cluster voltage deviation errors, which are illustrated in Fig.~\ref{fig:volterrorsIters}. These results clearly show that despite subjecting the optimal Kron reduction to a wide range of operating conditions, the worst-case intra-cluster voltage deviation errors are still very small across all super nodes and loading conditions (i.e., all super node clusters deviate from their corresponding reduced nodes by less than 0.0065pu). The fact that errors do not increase away from known input data scenarios $(V,I)$ (which are at either end in Fig.~\ref{fig:volterrorsIters}) may seem surprising. However, AC load flows are nonlinear, the optimal Kron-reduction minimizes the worst-case voltage errors, and the two load scenarios were low- and high-load conditions. This means that away from high-load conditions (which was in our initial set of data),  the voltages at each node will become closer to 1.00pu and, thus, closer to each other, which reduces voltage deviation errors.  Thus, including high net-load demand profiles to generate initial input data that has large voltage deviations may help find an optimal network reduction that captures the full system behavior accurately. In addition, the structure-preserving nature of the optimal Kron reduction appears quite valuable to represent a wide range of operating conditions. 

Lastly, to understand the effects of constraining the complexity at each iteration, we explored different upper bounds, $\beta = \{0.10, 0.25, 0.50, 0.75\}$. Then, we looked at the number of iterations required to achieve a converged optimal Kron-based network reduction, the level of the reduction, and the corresponding worst-case voltage errors. Results are summarized in Table~\ref{tab:compareBeta} and show that smaller bounds can reduce overall errors, but at the cost of the reduction itself. The best trade-off is $\beta=0.25$, with high level of reduction and voltage errors $<0.01$pu ($<10$\textit{mili}-pu or \textit{m}pu). 

\begin{table}[h]
   \caption{Different upper bounds on complexity ($\beta$)} 
   \label{tab:compareBeta}
   \small
   \centering
   \begin{tabular}{llcccc}
   \toprule\toprule
  \textbf{Network} & \textbf{Item}    & \textbf{$\beta = 0.10$}   & \textbf{$0.25$}   & \textbf{$0.50$} & \textbf{$0.75$} \\  
   \midrule 
   \multirow{3}{4em}{115-node radial} & Iterations (\#)       & 17                   & 8                         & 7                       &  7 \\
   &Reduction (\%)        & 75                   &  85                       & 83                      &  83.5 \\
   &Voltage err (\textit{m}pu)    & 3.0                & 6.5                   & 3.5                  & 5.0 \\
   \midrule 
   \multirow{3}{4em}{200-bus mesh} & Iterations (\#)       & 10                   & 9                         & 9                       &  8 \\
   &PQ Reduction (\%)          & 71.6                   & 75.3                         & 79.0                       &  81.5 \\
   &Voltage err (\textit{m}pu)       & 16                   & 17                         & 19                       &  20 \\

\bottomrule
   \end{tabular}
\end{table}

\subsection{200-node mesh network}
\begin{figure}[t]
\includegraphics[width=\columnwidth]{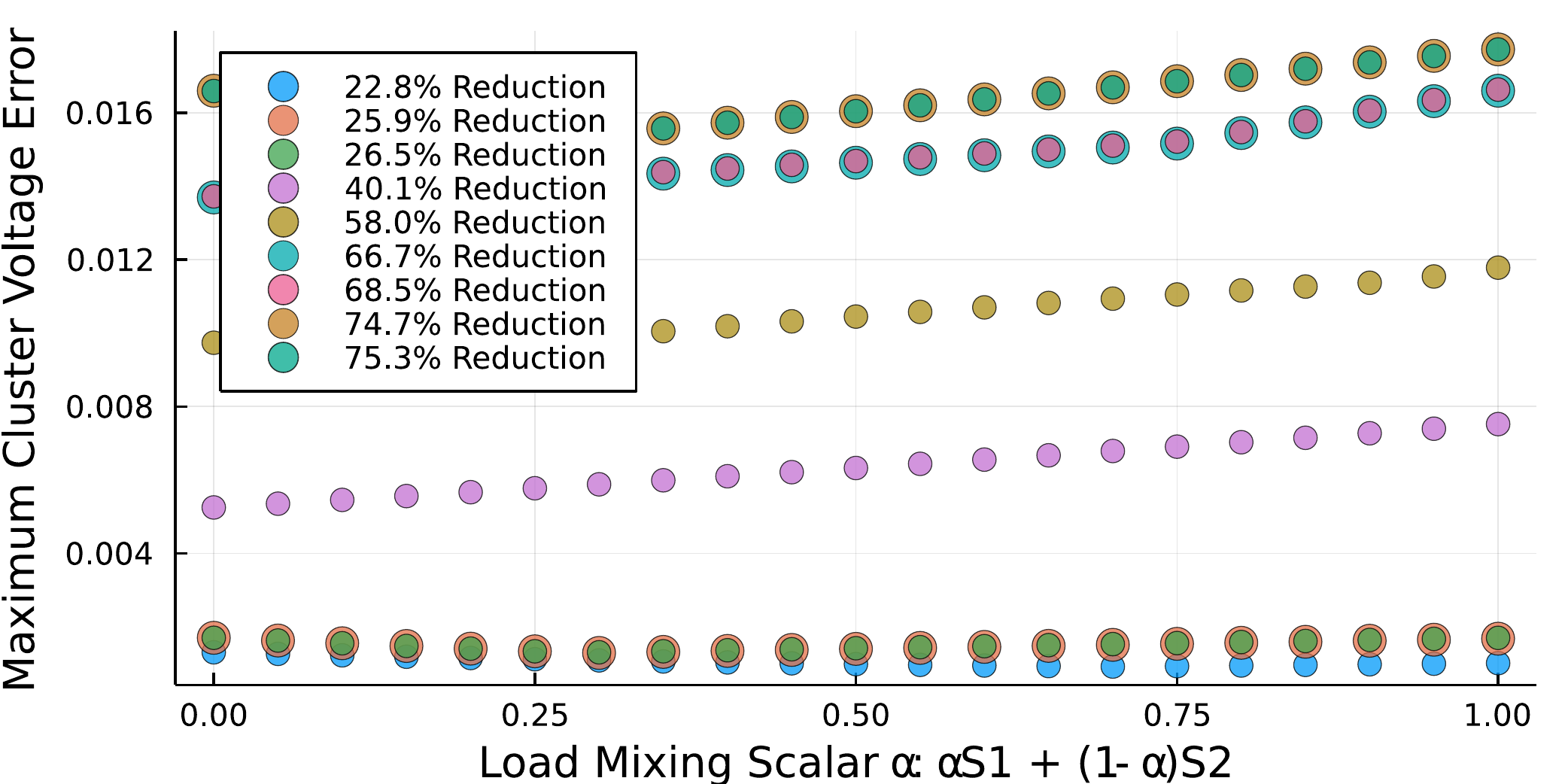}
\caption{Worst-case intra-cluster voltage errors for all iterate Kron-reduced versions of the meshed 200-node system (optimally reduced via nine iterations in Alg.~\ref{alg:one}). Reductions are given in terms of load bus reductions.}
\label{fig:volterrorsIters_mesh}
\centering
\end{figure}

In order to assess the performance of Opti-KRON in the context of a mesh network configuration, we also applied Algorithm~\ref{alg:one} to the 200 bus transmission system model from PGLib~\cite{Babaeinejadsarookolaee:2019}. The results associated with successive reductions (for $\alpha = 0.045$ and $\beta = 0.25$) are depicted in Fig.~\ref{fig:volterrorsIters_mesh}; the influence of $\beta$ is depicted in the bottom portion of Table \ref{tab:compareBeta}. When applying Algorithm~\ref{alg:one}, we added an additional constraint which prevented the reduction of any voltage-controlled generator buses (PV buses); only load buses (PQ buses) could be reduced. This assumption is consistent with the prevailing uses of Kron reduction in the literature~\cite{dorfler_kron_2013}.

Generally, Opti-KRON could achieve a fairly high level of load reduction in the mesh network, but when the level of reduction exceeded 50\%, the voltage error began to climb over 0.01pu. Furthermore, in the mesh network tests, the MILP solver required a significantly longer time to close the MIP gap (10s to 100s of seconds). Future work will seek to understand how Opti-KRON can be accelerated and improved when applied to mesh networks.

\section{Conclusion and Future work} \label{sec:end}

This paper develops a novel and efficient mixed-integer linear optimization-based methodology for generating structure-preserving network reductions of electric power networks. The MILP formulation enables trading off complexity (in the number of reduced nodes) and errors (in terms of worst-case voltage deviations across all super node clusters) and uses the network's graph Laplacian to restrict nodal eliminations to only include neighbors of chosen super nodes. By leveraging the efficient MILP formulation, an iterative scheme is employed to successively enhance the network reduction while ensuring that each iterate is a valid Kron reduction of the full network. Furthermore, simulation-based analysis is used to numerically explore the formulation and characterize and compare the optimal Kron reductions. The computational results illustrate that Opti-KRON can reduce full radial networks of more than 100 nodes by 25-90\% at optimality and within seconds. These optimal network reductions engender worst-case intra-cluster voltage magnitude deviations of less than 0.01pu. 

Future work will pursue a number of open questions resulting from discoveries herein. First, we will investigate the optimality gap of and compare conventional network reduction techniques against Opti-KRON. For example, while the iterative scheme is guaranteed to converge to a Kron-reduced network, we have not established global optimality guarantees at convergence. However, for radial networks, it may be possible to prove that the successive iterations will yield the globally optimal Kron reduced network~\cite{yuan_inverse_2021}. Furthermore, we are interested in using the optimal Kron-reduced networks in OPF problems and want to incorporate the corresponding worst-case intra-cluster voltage deviations to yield robust OPF formulations (e.g., via tightened voltage bounds) whose solutions guarantee admissibility in the underlying full network~\cite{almassalkhi_hierarchical_2020}. Similarly, solving the OPF on a reduced network will require a disaggregation policy to lift the optimal solution on the reduced network to the full network's (individual) resources, whose analysis is of interest. 

\appendix
\section{Reformulating into rectangular coordinates}
While the Opti-KRON formulation is stated in complex variables in~\eqref{eq:optiKron}, it was decomposed into purely real rectangular coordinates to be solved. Decomposing the admittance and impedance matrices into their real and imaginary parts, we have
\begin{align}
\left\langle Y_{b}\right\rangle=\left[\begin{array}{cc}
Y_{G} & -Y_{B}\\
Y_{B} & Y_{G}
\end{array}\right],\; \left\langle Z_{b}\right\rangle=\left[\begin{array}{cc}
Z_{G} & -Z_{B}\\
Z_{B} & Z_{G}
\end{array}\right].
\end{align}
To build the generalized Kron impedance, we extend the selection matrix into a block diagonal form:
\begin{align}
\left\langle Z_{K}\right\rangle =\left[\begin{array}{cc}
S & 0\\
0 & S
\end{array}\right]\left[\begin{array}{cc}
Z_{G} & -Z_{B}\\
Z_{B} & Z_{G}
\end{array}\right]\left[\begin{array}{cc}
S & 0\\
0 & S
\end{array}\right].
\end{align}
Likewise, for super node voltage selection and current aggregation, the following block diagonal expressions produce these quantities:
\begin{align}
\left[\!\!\begin{array}{c}
V_{s,r}\\
V_{s,i}
\end{array}\!\!\right]=\left[\!\begin{array}{cc}
S & 0\\
0 & S
\end{array}\!\right]\!\left[\!\begin{array}{c}
V_{r}\\
V_{i}
\end{array}\!\right],\left[\!\begin{array}{c}
I_{s,r}\\
V_{s,i}
\end{array}\!\right]\!=\!\left[\!\begin{array}{cc}
A & 0\\
0 & A
\end{array}\!\right]\!\left[\!\begin{array}{c}
I_{r}\\
I_{i}
\end{array}\!\right].\nonumber
\end{align}

\printbibliography

@ARTICLE{schneider_ieee_2017,
  author={Schneider, K. P. and Mather, B. A. and others}, %Pal, B. C. and Ten, C.-W. and Shirek, G. J. and Zhu, H. and Fuller, J. C. and Pereira, J. L. R. and Ochoa, L. F. and de Araujo, L. R. and Dugan, R. C. and Matthias, S. and Paudyal, S. and McDermott, T. E. and Kersting, W.}

@ARTICLE{Babaeinejadsarookolaee:2019,
       author = {{Babaeinejadsarookolaee}, Sogol and {Birchfield}, Adam and others},
        title = {The Power Grid Library for Benchmarking {AC} Optimal Power Flow Algorithms},
      journal = {arXiv:1908.02788},
         year = 2019,
        month = aug,
          eid = {arXiv:1908.02788},
        pages = {},
archivePrefix = {arXiv},
       eprint = {1908.02788},
 primaryClass = {math.OC},
       adsurl = {https://ui.adsabs.harvard.edu/abs/2019arXiv190802788B}
}

@ARTICLE{chevalier_accelerate_2021,
  author={Chevalier, Samuel and Schenato, Luca and Daniel, Luca},
  journal={IEEE Transactions on Power Systems}, 
  title={Accelerated Probabilistic Power Flow in Electrical Distribution Networks via Model Order Reduction and Neumann Series Expansion}, 
  year={2022},
  volume={37},
  number={3},
  pages={2151-2163},
  doi={10.1109/TPWRS.2021.3120911}}

@ARTICLE{dongchan_robust_2021,
  author={Lee, Dongchan and Turitsyn, Konstantin and Molzahn, Daniel K. and Roald, Line A.},
  journal={IEEE Transactions on Power Systems}, 
  title={{Robust AC Optimal Power Flow with Robust Convex Restriction}}, 
  year={2021},
  volume={36},
  number={6},
  pages={4953-4966},
  doi={10.1109/TPWRS.2021.3075925}}

@article{stott_dc_2009,
	title = {{DC} {Power} {Flow} {Revisited}},
	volume = {24},
	year = {2009},
	url = {http://ieeexplore.ieee.org/document/4956966/},
	doi = {10.1109/TPWRS.2009.2021235},
	number = {3},
	journal = {IEEE Transactions on Power Systems},
	author = {Stott, B and Jardim, J and Alsac, O},
	pages = {1290--1300},
}

@article{almassalkhi_hierarchical_2020,
	title = {Hierarchical, {Grid}-{Aware}, and {Economically} {Optimal} {Coordination} of {Distributed} {Energy} {Resources} in {Realistic} {Distribution} {Systems}},
	volume = {13},
	issn = {1996-1073},
	url = {https://www.mdpi.com/1996-1073/13/23/6399},
	doi = {10.3390/en13236399},
	abstract = {Renewable portfolio standards are targeting high levels of variable solar photovoltaics (PV) in electric distribution systems, which makes reliability more challenging to maintain for distribution system operators (DSOs). Distributed energy resources (DERs), including smart, connected appliances and PV inverters, represent responsive grid resources that can provide ﬂexibility to support the DSO in actively managing their networks to facilitate reliability under extreme levels of solar PV. This ﬂexibility can also be used to optimize system operations with respect to economic signals from wholesale energy and ancillary service markets. Here, we present a novel hierarchical scheme that actively controls behind-the-meter DERs to reliably manage each unbalanced distribution feeder and exploits the available ﬂexibility to ensure reliable operation and economically optimizes the entire distribution network. Each layer of the scheme employs advanced optimization methods at different timescales to ensure that the system operates within both grid and device limits. The hierarchy is validated in a large-scale realistic simulation based on data from the industry. Simulation results show that coordination of ﬂexibility improves both system reliability and economics, and enables greater penetration of solar PV. Discussion is also provided on the practical viability of the required communications and controls to implement the presented scheme within a large DSO.},
	language = {en},
	number = {23},
	urldate = {2022-01-19},
	journal = {Energies},
	author = {Almassalkhi, Mads and Brahma, Sarnaduti and Nazir, Nawaf and others},% Ossareh, Hamid and Racherla, Pavan and Kundu, Soumya and Nandanoori, Sai Pushpak and Ramachandran, Thiagarajan and Singhal, Ankit and Gayme, Dennice and Ji, Chengda and Mallada, Enrique and Shen, Yue and You, Pengcheng and Anand, Dhananjay}

@article{oh_new_2010,
	title = {A {New} {Network} {Reduction} {Methodology} for {Power} {System} {Planning} {Studies}},
	volume = {25},
	issn = {0885-8950, 1558-0679},
	url = {http://ieeexplore.ieee.org/document/5357378/},
	doi = {10.1109/TPWRS.2009.2036183},
	language = {en},
	number = {2},
	urldate = {2022-03-30},
	journal = {IEEE Transactions on Power Systems},
	author = {Oh, HyungSeon},
	month = may,
	year = {2010},
	pages = {677--684},
}

@article{shi_novel_2015,
	title = {A {Novel} {Bus}-{Aggregation}-{Based} {Structure}-{Preserving} {Power} {System} {Equivalent}},
	volume = {30},
	issn = {0885-8950, 1558-0679},
	url = {http://ieeexplore.ieee.org/document/6917217/},
	doi = {10.1109/TPWRS.2014.2359447},
	language = {en},
	number = {4},
	urldate = {2022-03-30},
	journal = {IEEE Transactions on Power Systems},
	author = {Shi, Di and Tylavsky, Daniel J.},
	month = jul,
	year = {2015},
	pages = {1977--1986},
}

@phdthesis{ploussard_efficient_2019,
	title = {Efficient reduction techniques for a large-scale {Transmission} {Expansion} {Planning} problem},
	language = {en},
	author = {Ploussard, Quentin},
	year = {2019},
}

@article{squires_economic_1960,
	title = {Economic {Dispatch} of {Generation} {Directly} {Rrom} {Power} {System} {Voltages} and {Admittances}},
	volume = {79},
	issn = {0097-2460},
	url = {http://ieeexplore.ieee.org/document/4500947/},
	doi = {10.1109/AIEEPAS.1960.4500947},
	language = {en},
	number = {3},
	urldate = {2022-03-30},
	journal = {Transactions of the American Institute of Electrical Engineers. Part III: Power Apparatus and Systems},
	author = {Squires, R. B.},
	month = apr,
	year = {1960},
	pages = {1235--1244},
}

@article{fortenbacher_transmission_2018,
	title = {Transmission {Network} {Reduction} {Method} using {Nonlinear} {Optimization}},
	url = {http://arxiv.org/abs/1711.01079},
	doi = {10.23919/PSCC.2018.8442974},
	language = {en},
	urldate = {2022-03-30},
	journal = {2018 Power Systems Computation Conference (PSCC)},
	author = {Fortenbacher, Philipp and Demiray, Turhan and Schaffner, Christian},
	month = jun,
	year = {2018},
	note = {arXiv: 1711.01079},
	pages = {1--7},
}

@article{sistermanns_feature-_2019,
	title = {Feature- and {Structure}-{Preserving} {Network} {Reduction} for {Large}-{Scale} {Transmission} {Grids}},
	url = {http://arxiv.org/abs/1903.11590},
	doi = {10.1109/PTC.2019.8810704},
	language = {en},
	urldate = {2022-03-30},
	journal = {2019 IEEE Milan PowerTech},
	author = {Sistermanns, Julia and Hotz, Matthias and Hewes, Dominic and Witzmann, Rolf and Utschick, Wolfgang},
	month = jun,
	year = {2019},
	note = {arXiv: 1903.11590},
	pages = {1--6},
}

@article{yuan_inverse_2021,
	title = {Inverse {Power} {Flow} {Problem}},
	url = {http://arxiv.org/abs/1610.06631},
	language = {en},
	urldate = {2022-03-30},
	journal = {arXiv:1610.06631 [cs, eess, math]},
	author = {Yuan, Ye and Low, Steven and Ardakanian, Omid and Tomlin, Claire},
	month = oct,
	year = {2021},
	note = {arXiv: 1610.06631},
}

@article{ploussard_efficient_2018,
	title = {An {Efficient} {Network} {Reduction} {Method} for {Transmission} {Expansion} {Planning} {Using} {Multicut} {Problem} and {Kron} {Reduction}},
	volume = {33},
	issn = {0885-8950, 1558-0679},
	url = {https://ieeexplore.ieee.org/document/8370048/},
	doi = {10.1109/TPWRS.2018.2842301},
	language = {en},
	number = {6},
	urldate = {2022-03-30},
	journal = {IEEE Transactions on Power Systems},
	author = {Ploussard, Quentin and Olmos, Luis and Ramos, Andres},
	month = nov,
	year = {2018},
	pages = {6120--6130},
}

@article{rogers_aggregation_1991,
	title = {Aggregation and {Disaggregation} {Techniques} and {Methodology} in {Optimization}},
	volume = {39},
	issn = {0030-364X, 1526-5463},
	url = {http://pubsonline.informs.org/doi/abs/10.1287/opre.39.4.553},
	doi = {10.1287/opre.39.4.553},
	language = {en},
	number = {4},
	urldate = {2022-03-30},
	journal = {Operations Research},
	author = {Rogers, David F. and Plante, Robert D. and Wong, Richard T. and Evans, James R.},
	month = aug,
	year = {1991},
	pages = {553--582},
}

@article{kokotovic_singular_1976,
	title = {Singular perturbations and order reduction in control theory—an overview},
	volume = {12},
	number = {2},
	journal = {Automatica},
	author = {Kokotovic, Petar V and O'Malley Jr, Robert E and Sannuti, Peddapullaiah},
	year = {1976},
	note = {Publisher: Elsevier},
	pages = {123--132},
}

@article{carpentier_optimal_1979,
	title = {Optimal power flows},
	volume = {1},
	number = {1},
	journal = {International Journal of Electrical Power \& Energy Systems},
	author = {Carpentier, J},
	year = {1979},
	note = {Publisher: Elsevier},
	pages = {3--15},
}

@article{dommel_optimal_1968,
	title = {Optimal {Power} {Flow} {Solutions}},
	volume = {PAS-87},
	doi = {10.1109/TPAS.1968.292150},
	number = {10},
	journal = {IEEE Transactions on Power Apparatus and Systems},
	author = {Dommel, Hermann W. and Tinney, William F.},
	year = {1968},
	pages = {1866--1876},
}

@article{dorfler_kron_2013,
	title = {Kron {Reduction} of {Graphs} {With} {Applications} to {Electrical} {Networks}},
	volume = {60},
	doi = {10.1109/TCSI.2012.2215780},
	number = {1},
	journal = {IEEE Transactions on Circuits and Systems I: Regular Papers},
	author = {Dorfler, Florian and Bullo, Francesco},
	year = {2013},
	pages = {150--163},
}

@article{molzahn_survey_2019,
	title = {A {Survey} of {Relaxations} and {Approximations} of the {Power} {Flow} {Equations}},
	volume = {4},
	issn = {2332-6557, 2332-6565},
	url = {http://www.nowpublishers.com/article/Details/EES-012},
	doi = {10.1561/3100000012},
	language = {en},
	number = {1-2},
	urldate = {2022-03-30},
	journal = {Foundations and Trends® in Electric Energy Systems},
	author = {Molzahn, Daniel K. and Hiskens, Ian A.},
	year = {2019},
	pages = {1--221},
}

@article{low_convex_2014,
	title = {Convex {Relaxation} of {Optimal} {Power} {Flow}—{Part} {I}: {Formulations} and {Equivalence}},
	volume = {1},
	issn = {2325-5870},
	shorttitle = {Convex {Relaxation} of {Optimal} {Power} {Flow}—{Part} {I}},
	url = {http://ieeexplore.ieee.org/document/6756976/},
	doi = {10.1109/TCNS.2014.2309732},
	language = {en},
	number = {1},
	urldate = {2022-03-30},
	journal = {IEEE Transactions on Control of Network Systems},
	author = {Low, Steven H.},
	month = mar,
	year = {2014},
	pages = {15--27},
}

@article{dallanese_optimal_2018,
	title = {Optimal {Power} {Flow} {Pursuit}},
	volume = {9},
	issn = {1949-3053, 1949-3061},
	url = {http://ieeexplore.ieee.org/document/7480375/},
	doi = {10.1109/TSG.2016.2571982},
	language = {en},
	number = {2},
	urldate = {2022-03-30},
	journal = {IEEE Transactions on Smart Grid},
	author = {Dall'Anese, Emiliano and Simonetto, Andrea},
	month = mar,
	year = {2018},
	pages = {942--952},
}
\addtolength{\textheight}{-12cm}   

\end{document}